\documentclass{article}
\usepackage[utf8]{inputenc}
\usepackage[english]{babel}
\usepackage[a4paper, total={6in, 8in}]{geometry}
\usepackage{comment}
\usepackage{dcolumn}
\usepackage{placeins}
\usepackage{rotating}

\usepackage[backend=biber,style=ieee]{biblatex}
\providecommand{\keywords}[1]
{
  \small	
  \textbf{\textit{Keywords---}} #1
}

\title{Understanding Factors that Influence Upskilling}
\author{
Eduardo Laguna Muggenburg\\
  \texttt{emuggenburg@fb.com}
  \and
   Monica Bhole\\
  \texttt{mbhole@fb.com}
  \and
  Michael Meaney\\
  \texttt{meaney@fb.com}
  
}
\date{\today}

\usepackage{graphicx}
\usepackage[flushleft]{threeparttable}

\usepackage{hyperref}

\hypersetup{
    colorlinks=true,
    linkcolor=blue,
    filecolor=magenta,      
    urlcolor=cyan,
    pdftitle={Sharelatex Example},
    bookmarks=true,
    pdfpagemode=FullScreen,
    }
    
\begin{document}





\maketitle

\begin{abstract}

We investigate the motivation and means through which individuals expand their skill-set by analyzing a survey of applicants from the Facebook Jobs product. Individuals who report being influenced by their networks or local economy are over 29\% more likely to have a postsecondary degree, but peer effects still exist among those who do not acknowledge such influences. Users with postsecondary degrees are more likely to upskill in general, by continuing coursework or applying to higher-skill jobs, though the latter is more common among users across all education backgrounds. These findings indicate that policies aimed at connecting individuals with different educational backgrounds can encourage upskilling. Policies that encourage users to enroll in coursework may not be as effective among individuals with a high school degree or less. Instead, connecting such individuals to opportunities that value skills acquired outside of a formal education, and allow for on-the-job training, may be more effective. 

\end{abstract}


\section{Introduction}
\label{sec:intro}



Skills-biased technical change has led to sustained declines in wages for less educated workers \cite{NBERw25588}, with COVID-19 further accelerating these trends \cite{autor2020nature}. Recent policy discussions have centered on how workforce development and higher educational infrastructure could help workers adapt to these changes \cite{escobari2019realism}. While the literature on skill-biased technical change \cite{david2020extending} and the returns to education \cite{oreopoulos2013making} is more developed, research investigating the motivation to pursue further education as an adult, and the perceived value in doing so, is relatively sparse. Better understanding these issues is essential to developing policies that can help those most adversely affected by the growing wage gap between skilled and unskilled workers. We investigate these issues by analyzing a sample of Facebook Jobs applicants in the United States (US) to understand factors that may influence the decision to pursue a higher education in the first place, the perceived value of education to employment, and varying approaches to upskilling. 

First, we consider whether users in our sample were influenced by their friends or family (network influences), or by the local economy (local influence), when making decisions about postsecondary education. Existing literature suggests network influences, including exposure to others with a postsecondary degree on social media, increases the likelihood of completing postsecondary education \cite{cataldi2018first, palardy2013high, sacerdote2011peer}. Additionally, local labor market information can positively influence enrollment behavior and major selection in community colleges \cite{foote2020effect, baker2018effect}. We find that individuals who acknowledge these influences are over 29\% more likely to have a postsecondary degree. Among users less than 30 years old, those with a larger share of college Facebook friends while in high school (HS) were over 40\% more likely to have a postsecondary degree, confirming that peer effects exist even among those who do not acknowledge them. 

Given that peer effects appear to influence the decision to pursue a postsecondary degree, we next consider the perceived return to education. The existing literature in this area is less developed. Surveys generally reveal that that more highly educated populations are more likely to engage in learning throughout their lives \cite{horrigan2016lifelong}, and hold a higher perception of education’s value to their job \cite{morin2014rising}.  We find that those with a postsecondary degree are over 30\% more likely to indicate that the skills from their education are relevant to their latest job. When we ask respondents whether they have recently enrolled in any courses, those who believe their education provided skills relevant to their latest job, and those with a postsecondary degree, were over 37\% more likely to have continued their education in the last year than those with a HS degree or less. 

Formal education is not the only means through which individuals can improve their labor market outcomes. Not all jobs held by those with a college degree require such credentials and those who do not believe their education relevant may try to upskill on the job. To understand this, we asked applicants whether they have skills relevant to the jobs to which they applied, and whether these skills were acquired through a formal education. Those with a two-year or Bachelor’s degree were 34\% less likely to indicate they applied to jobs unrelated to their skill-set (and were thus looking to expand their skill-set). Those with a higher degree were more likely to apply to jobs that required skills acquired through their formal education.

Finally, we investigate how much individuals participate in 'observable upskilling' by studying applications sent on the platform after our survey. To achieve this, we asked users for their latest job title and to self-categorize their latest job into one of eight International Labor Organization (ILO) skill groups \cite{ILOclassification}. We then apply the same categorization to the jobs to which users apply and evaluate whether applicants apply to jobs in higher skill groups. Those with a higher education are more likely to upskill through their applications, as are those who are employed and looking for work. Whereas those with a Bachelor’s degree were 37\% (8.9 percentage points) more likely to have taken courses than those with a HS degree, they were only 15\% (5.5 percentage points) more likely to have 'observably upskilled' through their applications, a result that is not statistically significant. These findings indicate that the more educated are generally more likely to upskill, through education or their job applications, but the differences are much smaller when we consider upskilling through applications. 

Our results are particularly relevant to explore in the context of the COVID-19 economy, where massive job losses have disproportionately affected lower-wage workers, in doing so accelerating long observed trends of skills-biased technical change, with present workforce infrastructure ill-equipped to help \cite{NBERw25588, autor2020nature, david2020extending, escobari2019realism}. 

In section \ref{sec:Data} we briefly describe our data, and our analysis of the relationship between the influence of peer networks and the local economy are in section \ref{sec:jes}. Section \ref{sec:upskilling} studies the ways in which individuals in our sample upskill, and section \ref{sec:conclusion} concludes.

\section{Data}
\label{sec:Data}




We surveyed a random sample of Facebook Jobs applicants in the US who submitted an application through the product between January 23 and February 15, 2020. Respondents were recruited through the Facebook Newsfeed and 10,896 individuals completed the survey. We focus on respondents who answered questions about their educational attainment and their employment status, leading to a final sample of 9,568 users. To adjust for non-response bias, we use inverse propensity score weighting to build survey weights for our sample and make it representative of the users in our primary sample \cite{valliant2013}. While users in our sample applied to a job on Facebook in the beginning of 2020, the survey was conducted from April 30th to May 19th. The users we surveyed applied before the US was affected by shelter-in-place policies due to COVID-19.


Table \ref{tab:sumstats} summarizes individuals in our survey. The average applicant was 38 years old, had been using Facebook for 8.3 years, and had 821 Facebook friends. Based on survey responses, 6.43\% had no degree, 52.1\% had a HS degree equivalent, 27.6\% had a two-year postsecondary degree and 13.9\% had a Bachelor's or higher. We use this measure of educational attainment throughout the rest of this paper. In our sample, 31.5\% of individuals were unemployed and another 29.6\% were employed but looking for jobs when the survey was conducted. While 28\% of the US population has a high school degree equivalent, our sample over-indexes on this population \cite{eduattainmentUSA2019}. This is an important population to understand, however, as they are likely to be affected by skill-biased technical change. While our sample consists of users who applied to a job in January or February of 2020, by May 30.3\% of these applicants were employed and not actively searching for work.

\begin{table}[h]
    \caption{Summary Statistics}\label{tab:sumstats}
    \centering
    \begin{tabular}{lccc}
    \hline \hline
         & Mean & S.D.   \\
         \hline
        Age & 38 & 12.81 \\
        \# Friends & 821.71 &  986.24 \\
        Years on FB & 8.29 & 3.78 \\
        No Degree & 6.43 & 24.52  \\
        HS or GED & 52.11 & 49.96  \\
        2 year postsecondary degree & 27.56 & 44.68  \\
        4 year postsecondary degree or higher & 13.90 & 34.59  \\
        Unemployed (\%) & 31.56 & 46.47  \\
        Employed and looking (\%) & 29.58 & 45.64  \\
        Continued Edu (last 12 mos) & 0.24 & 0.42 \\
        Applied to job (3 months post survey) & 0.17 & 0.37 \\
        Upskill at all & 0.36 & 0.48\\
        N & 9568 & \\
        \hline
        \hline
    \end{tabular}
\end{table}
\FloatBarrier

Table \ref{tab:sumstats} summarizes upskilling and post-survey application behavior of individuals in our sample. Overall 24\% of users had taken courses in the previous year, one method of upskilling, and 17\% had applied to a job within the 3 months following the survey. By comparing the skill level associated with users' latest job, as of our survey, to the skill level of the job users most recently applied to we can understand the characteristics of individual who participate in 'observable upskilling,' meaning they applied to higher-skill jobs. In addition to continuing education, and 'observable upskilling,' we asked respondents how the jobs they applied to in January related to their skills and allowed users to indicate whether their skills were unrelated, but they'd like to expand their skill-set. This, continuing education, and 'observable upskilling' are all methods of upskilling we can measure. Overall, 36\% of users upskilled through one of these three methods. 

\begin{figure}[ht]
    \centering
    \includegraphics[width=.8\textwidth]{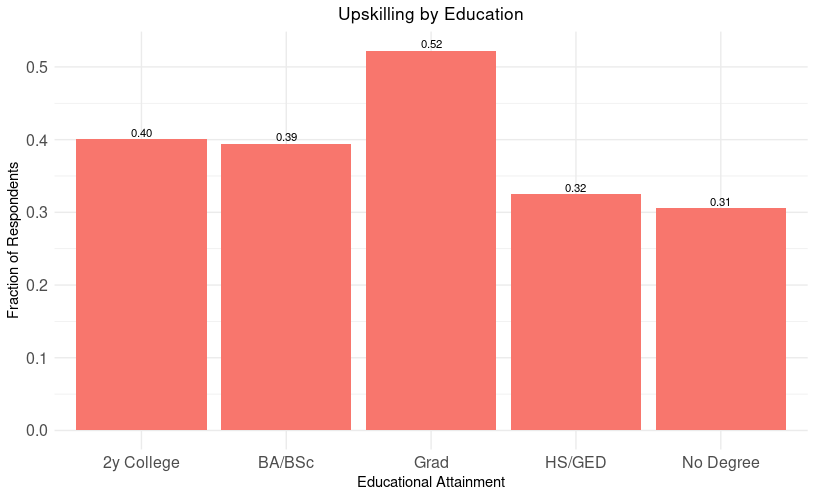}
    \caption{Upskilling by Education Status}
    \label{fig:upskill_ed_sumamry}
\end{figure}

While a large fraction of users appear to upskill in general, Figure \ref{fig:upskill_ed_sumamry} shows users with a postsecondary degree or higher are more likely to upskill via one of these methods. Overall, 52\% of those with a graduate degree and 40\% of those with a two-year or Bachelor's degree appeared to be upskilling. These percentages are slightly smaller among those with a HS degree or less. In the following sections we will compare methods of upskilling across users to understand which individuals are more likely to upskill through their applications versus additional education. 

\section{Influence of Networks on Educational Choice}
\label{sec:jes}
To understand how individuals' environment affects the decision to pursue postsecondary education we asked respondents whether their decision to pursue higher education was influenced by their peers or the local economy. Understanding whether the local job market affects an individual's decision to pursue higher education can have important implications if policymakers want to ensure that individuals are prepared for a changing economy. Similarly, understanding whether peers affect this decision can also inform initiatives to help individuals expand their skill-set.

\begin{figure}[ht]
    \centering
    \includegraphics[width=.8\textwidth]{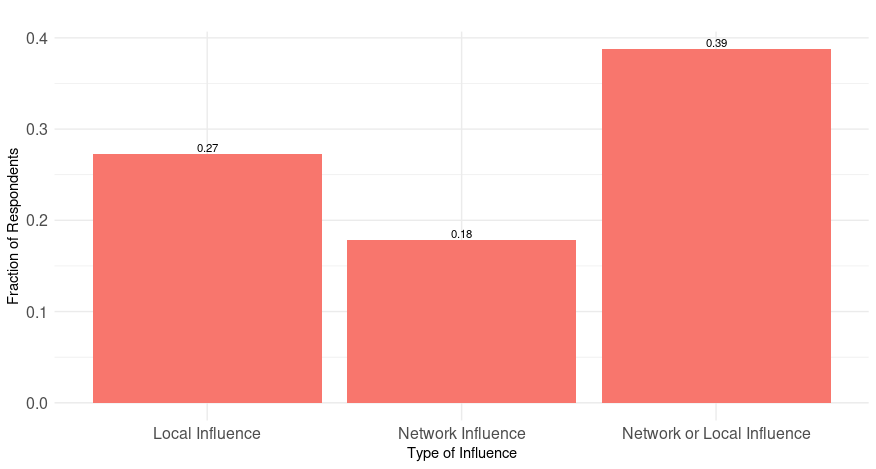}
    \caption{Summary of users who acknowledge local (left) or network (middle) influence on educational choice. The rightmost bar includes users who acknowledged the either local or network influences (or both).}
    \label{fig:local_friend_summary}
\end{figure}

Figure \ref{fig:local_friend_summary} summarizes the fraction of users who indicate that the decision to pursue education was influenced by their friends or family (network influence) and the fraction of users who indicate the decision to pursue education was influenced by the local economy when they made the choice (local influence). Users could select either or both option. Overall, 18\% indicate they were influenced by their network while 27\% were influenced by the local economy. This includes users who said both. Overall, 39\% of users acknowledge the influence of either option on their educational choice. 

In Table \ref{tab:table2} we consider the relationship between acknowledgment of a network or local influence and educational attainment by estimating linear probability models (LPM). Across all specifications we control for number of friends, gender, age, median income of the home zipcode as well as characteristics of educational opportunities in the county for the HS they specify on their profile. In specifications 1 and 2 we focus on the entire sample of users, and in specification 2 we include  controls for the share of high-skilled jobs, and the number of jobs, in the user's county when they were in HS (based on their profile information). Specification 2 indicates that users who acknowledge that they were influenced by their networks in HS are 12.1 percentage points (pp) more likely (29\% relative to the mean) to have a postsecondary degree. Users influenced by the local labor market were 15.5pp more likely  (37\% relative to the mean) to have a postsecondary degree than those who were not influenced. In addition to the 12.1 and 15.1 estimates, those influenced by both were an additional 7.4pp more likely to have a college education -- translating to over an 80\% higher probability of having a college education. Among those who acknowledge that they were influenced by their networks and surroundings, these factors are correlated with a higher probability of pursuing higher education. 


\begin{table}[!htbp] {\centering 
  \caption{Educational Attainment} 
  \label{tab:table2} 
\begin{tabular}{@{\extracolsep{5pt}}lD{.}{.}{-3} D{.}{.}{-3} D{.}{.}{-3} D{.}{.}{-3} } 
\\[-1.8ex]\hline 
\hline \\[-1.8ex] 
 & \multicolumn{4}{c}{\textit{Dependent variable:}} \\ 
\cline{2-5} 
\\[-1.8ex] & \multicolumn{4}{c}{Highest Degree Attained is College or More} \\ 
\\[-1.8ex] & \multicolumn{1}{c}{(1)} & \multicolumn{1}{c}{(2)} & \multicolumn{1}{c}{(3)} & \multicolumn{1}{c}{(4)}\\ 
\hline \\[-1.8ex] 
Network Influence  & 0.122^{***} & 0.121^{***} & 0.128^{***} & 0.078^{**} \\ 
  & (0.018) & (0.018) & (0.029) & (0.033) \\ 
  & & & & \\ 
Local Influence & 0.154^{***} & 0.155^{***} &  & 0.148^{***} \\ 
  & (0.013) & (0.017) &  & (0.034) \\ 
  & & & & \\ 
Share High-skill Jobs &  & -0.020 &  & -0.378^{*} \\ 
in HS county (HJ) &  & (0.110) &  & (0.208) \\ 
  & & & & \\ 
Above Median Friends &  &  & 0.105^{***} & 0.101^{***} \\ 
 in HS with College (FC) &  &  & (0.031) & (0.030) \\ 
  & & & & \\ 

Network Influence & 0.074^{**} & 0.074^{**} &  & 0.092^{*} \\ 
 x Local Influence   & (0.029) & (0.029) &  & (0.048) \\ 
  & & & & \\ 
 Local Influence  x HJ &  & -0.004 &  & -0.120 \\ 
  &  & (0.066) &  & (0.125) \\ 
  & & & & \\ 
Network Influence x FC &  &  & 0.123^{**} & 0.141^{***} \\ 
  &  &  & (0.050) & (0.049) \\ 
  & & & & \\ 
\hline \\[-1.8ex] 
Dep. Mean & 0.41 & 0.41 & 0.30 & 0.30 \\ 
Sample & \multicolumn{1}{c}{Full} & \multicolumn{1}{c}{Full} & \multicolumn{1}{c}{Age $\leq$ 30}  & \multicolumn{1}{c}{Age $\leq$ 30}\\
Observations & \multicolumn{1}{c}{9,568} & \multicolumn{1}{c}{9,568} & \multicolumn{1}{c}{2,581} & \multicolumn{1}{c}{2,581} \\ 
Log Likelihood & \multicolumn{1}{c}{-6,562.490} & \multicolumn{1}{c}{-6,548.550} & \multicolumn{1}{c}{-1,589.764} & \multicolumn{1}{c}{-1,548.275} \\ 
Akaike Inf. Crit. & \multicolumn{1}{c}{13,146.980} & \multicolumn{1}{c}{13,167.100} & \multicolumn{1}{c}{3,227.528} & \multicolumn{1}{c}{3,172.550} \\ 
\hline 
\hline 
\end{tabular} }
\begin{tablenotes}
\item {Note: Each column present selected coefficients of a different Linear Probability Model with the dependent variable specified on top. All specifications include controls for number of friends, gender, age, median income of the home zipcode as well as characteristics of educational opportunities in their high school county. Robust standard errors presented in parentheses. $^{*}$p$<$0.1; $^{**}$p$<$0.05; $^{***}$p$<$0.01}
\end{tablenotes}  
\end{table}

One issue with these results is that users who acknowledge a local or network influence may be quite different from the average respondent. To understand this, we focus on users who are 30 years or younger in our survey in specifications 3 - 4, as we have a better understanding of these users' network when they were in HS. For these users we observe Facebook connections when they were 17 years old (if they were on Facebook), and whether these friends included a college network in their profile at the time of this friendship.
We construct the share of the user's friends that had attended college when the user was in HS. For the median HS user, 16\% of their friends had attended college. We incorporate this into our model as a binary indicator for users with a share of "college" friends above the median. Specifications 3 and 4 confirm that in this sample users who said they were influenced by their networks and local markets were still much more likely to have a postsecondary degree. In specifications 3 and 5 we include our indicator for whether the user had an above-median share of college friends in HS and interact this with the indicator for whether the user said they were influenced by their friends. Users who had college friends when they were in high school were 10pp (33\%) more likely to have a college degree. Users who affirm that they were influenced by their peers are even more likely to have a postsecondary education themselves. These results show that many users are influenced by their networks even if they don't realize this. While not necessarily surprising, given the peer effects literature, the table shows that peer effects exist among individuals more generally -- not just among those who acknowledge these effects. One implication of this is that connecting individuals to others pursuing a higher education can positively influence one's decision to pursue a higher education. 

\FloatBarrier 

\section{Returns to an Education and Methods of Upskilling}
\label{sec:upskilling}
The decision to pursue a higher education often occurs at a relatively young age when the return to that education is not always clear. Throughout life we can continue to expand our skill-set by pursuing more education or by learning on-the-job. In this section we study the perceived relationship between education and the most recent job users held. We also study the extent to which users in our survey expand their skill-set by applying to higher-skill jobs or by taking classes in the past 12 months. 

Understanding these patterns is important for a number of reasons. First, this can help us understand why one's peers may influence the decision to pursue higher education. Second, evaluating the perceived return to an education, and the desire to learn through courses versus on the job, can help us evaluate the potential efficacy of educational versus hands-on retraining programs. If most individuals do not believe that higher education is relevant to their career, or are not interested in enrolling in courses, encouraging individuals to pursue higher education is unlikely to help those whose jobs are most at risk from technological change. If instead individuals try to expand their skills by applying to higher-skilled jobs, or indicate that they'd like to learn on the job, then apprenticeships, rotational programs, or programs that incentivize employers to train inexperienced employees may be more effective. 

To understand how users perceive their own education, we asked respondents whether their education or outside skills were more relevant to their current job and to the job that users applied to in January. In Specification 1 of Table \ref{tab:value_school} we focus on users who were employed during the survey and consider whether skills acquired through education were relevant to these users' current jobs by estimating a LPM. Overall, 34\% of these users indicated that skills acquired from their education were relevant to their current job. Those with a postsecondary degree are over 30\% more likely to indicate that the skills from their education are relevant to their latest job. More educated users were more likely to indicate that their education is useful; for example, those with a graduate degree were 18pp (50\%) more likely than those with a two-year degree to indicate that the skills acquired through their education is useful for their job. Users without a HS degree did not respond very differently from users with less than a HS degree (the baseline group). Related, users who took some coursework in the last year were 12.6pp (37\%) more likely to indicate that their education is useful to their current job. Users who were employed but looking for new opportunities were 4.8pp (14\%) less likely to indicate their education skills were relevant to their current job. These results are statistically significant at the 5\% level.


Specifications 2 - 5 consider whether applicants' skills were relevant to the jobs to which users applied in January, and the source of these skills (for users who remembered applying). Each column presents the results from a LPM for different outcomes. Overall, 35\% of users indicate that skills acquired outside of school were most relevant to the jobs they applied to, 10\% of users indicated their education was most relevant and 24\% of users said both were important. Another 22\% of users indicated that neither were relevant, but they were looking to expand their skill-set. Those with a two-year or Bachelor's degree were 7.5pp (34\%) less likely to indicate they applied to jobs unrelated to their skill-set (and were thus looking to expand their skill-set). Users with more education were more likely to apply to jobs relevant to their education. Users with a Bachelor's degree were 8.5pp (25\%) less likely than those without a degree to apply to jobs for which their out-of-schools skills are most relevant (specification 4), and those with a graduate degree were 41\% less likely to apply such jobs. These differences are statistically significant at the 5\% level.




\begin{table}[h]{ \centering 
  \caption{Current Skill-set and Job Applications} 
  \label{tab:value_school} 
\scalebox{.75}{
\begin{tabular}{@{\extracolsep{5pt}}lD{.}{.}{-3} D{.}{.}{-3} D{.}{.}{-3} D{.}{.}{-3} D{.}{.}{-3} } 
\\[-1.8ex]\hline 
\hline \\[-1.8ex] 
 & \multicolumn{5}{c}{\textit{Dependent variable:}} \\ 
\cline{2-6} 
\\[-1.8ex] & \multicolumn{1}{c}{School Skills} & \multicolumn{1}{c}{Applied to Job} & \multicolumn{3}{c}{Applied to Job Related to} \\ 
\\[-1.8ex] & \multicolumn{1}{c}{Important for Current Job} & \multicolumn{1}{c}{to Expand Skill-set} & \multicolumn{1}{c}{School Skills} & \multicolumn{1}{c}{Out-of-School Skills} & \multicolumn{1}{c}{Both Skills} \\ 
\\[-1.8ex] & \multicolumn{1}{c}{(1)} & \multicolumn{1}{c}{(2)} & \multicolumn{1}{c}{(3)} & \multicolumn{1}{c}{(4)} & \multicolumn{1}{c}{(5)}\\ 
\hline \\[-1.8ex] 
HS/GED & 0.041 & -0.035 & 0.027^{**} & 0.031 & 0.028 \\ 
  & (0.029) & (0.026) & (0.012) & (0.027) & (0.021) \\ 
  & & & & & \\ 
2y College & 0.147^{***} & -0.075^{***} & 0.077^{***} & -0.039 & 0.103^{***} \\ 
  & (0.031) & (0.026) & (0.014) & (0.028) & (0.023) \\ 
  & & & & & \\ 
BA/BSc & 0.186^{***} & -0.076^{***} & 0.072^{***} & -0.085^{***} & 0.140^{***} \\ 
  & (0.036) & (0.029) & (0.018) & (0.031) & (0.028) \\ 
  & & & & & \\ 
Grad & 0.308^{***} & -0.045 & 0.107^{***} & -0.144^{***} & 0.099^{***} \\ 
  & (0.044) & (0.036) & (0.026) & (0.037) & (0.035) \\ 
  & & & & & \\ 
Employed Looking & -0.048^{***} \\ 
  & (0.013)\\
    & & & & & \\ 
Classes in last 12 mos & 0.126^{***} & -0.016 & 0.033^{***} & -0.017 & 0.050^{***} \\ 
  & (0.016) & (0.012) & (0.010) & (0.014) & (0.013) \\ 
  & & & & & \\ 
\hline \\[-1.8ex] 
Dep. Mean & 0.34 & 0.22 & 0.10 & 0.35 & 0.24 \\ 
Sample & \multicolumn{1}{c}{Employed} & \multicolumn{1}{c}{Remember Jan App} & \multicolumn{1}{c}{Remember Jan App} & \multicolumn{1}{c}{Remember Jan App} & \multicolumn{1}{c}{Remember Jan App} \\
Observations & \multicolumn{1}{c}{5,469} & \multicolumn{1}{c}{7,068} & \multicolumn{1}{c}{7,068} & \multicolumn{1}{c}{7,068} & \multicolumn{1}{c}{7,068} \\ 
Log Likelihood & \multicolumn{1}{c}{-3,561.722} & \multicolumn{1}{c}{-3,951.073} & \multicolumn{1}{c}{-1,526.322} & \multicolumn{1}{c}{-4,883.314} & \multicolumn{1}{c}{-4,136.127} \\ 
Akaike Inf. Crit. & \multicolumn{1}{c}{7,177.445} & \multicolumn{1}{c}{7,960.145} & \multicolumn{1}{c}{3,110.643} & \multicolumn{1}{c}{9,824.628} & \multicolumn{1}{c}{8,330.253} \\ 
\hline 
\hline
\end{tabular} }}
\scriptsize\emph{Note:} Note: Each column present selected coefficients of a different Linear Probability Model with the dependent variable specified on top. All specifications include controls for number of friends, gender, age, median income of the home zipcode as well as characteristics of educational opportunities in their high school county. Columns 2 through 5 also control for individuals unemployment status. Robust standard errors presented in parentheses. $^{*}$p$<$0.1; $^{**}$p$<$0.05; $^{***}$p$<$0.01 \hfill
\end{table}

\FloatBarrier
Next we consider whether users in our sample were seeking to upskill, beyond applications sent in January. One method of upskilling is taking coursework in the past 12 months. We model the probability of taking coursework using a LPM in specification 1 of Table \ref{tab:upskilling_combined}. Overall, 24\% of users upskilled in the past 12 months by taking coursework. Those with a postsecondary degree were over 14.7pp (37\%) more likely to have continued their education in the last year than those with a high school degree or less. Users with a Bachelor's were 8.9 pp (37\%) more likely than those with a high school a degree to enroll in coursework and those with a graduate degree were 92\% more likely than those with a high school degree to upskill through coursework. 

We can also consider whether users apply to jobs at a higher skill level than their latest position by looking at users applying on the platform in the 3 months after our survey. Specification 2 of Table \ref{tab:upskilling_combined} shows that 17\% of users were still applying to jobs after the survey. The LPM in specification 2 shows that unemployed users were 3.1 pp (18\%) more likely than employed users who were not actively searching to apply to a job, while users who were employed and actively searching were 13\% more likely to apply. Interestingly, respondents who indicated they were impacted by COVID-19 were not more likely to apply during this period.

\begin{table}[hb]{\centering
  \caption{Up-skilling}\label{tab:upskilling_combined}
\begin{tabular}{@{\extracolsep{5pt}}lD{.}{.}{-3} D{.}{.}{-3} D{.}{.}{-3} } 
\\[-1.8ex]\hline 
\hline \\[-1.8ex] 
 & \multicolumn{3}{c}{Application Behavior} \\ 
\cline{2-4} 
\\[-1.8ex] & \multicolumn{1}{c}{Continued Education in} & \multicolumn{1}{c}{Is Still Applying for Jobs} & \multicolumn{1}{c}{Observable } \\ 
\\[-1.8ex] & \multicolumn{1}{c}{Previous 12 Months} & \multicolumn{1}{c}{After Survey} & \multicolumn{1}{c}{Up-skilling} \\ 
\\[-1.8ex] & \multicolumn{1}{c}{(1)} & \multicolumn{1}{c}{(2)} & \multicolumn{1}{c}{(3)}\\ 
\hline \\[-1.8ex] 
HS/GED & 0.026 & 0.009 & 0.099^{*} \\ 
  & (0.019) & (0.008) & (0.056) \\ 
  & & & \\ 
2y College & 0.128^{***} & 0.011 & 0.100^{*} \\ 
  & (0.021) & (0.009) & (0.059) \\ 
  & & & \\ 
BA/BSc & 0.115^{***} & 0.012 & 0.154^{**} \\ 
  & (0.025) & (0.010) & (0.069) \\ 
  & & & \\ 
Grad & 0.247^{***} & 0.013 & 0.183^{**} \\ 
  & (0.032) & (0.013) & (0.087) \\ 
  & & & \\ 
Unemployed & -0.014 & 0.031^{***} & 0.040 \\ 
  & (0.011) & (0.006) & (0.035) \\ 
  & & & \\ 
Employed Looking & 0.016 & 0.022^{***} & 0.068^{**} \\ 
  & (0.012) & (0.005) & (0.032) \\ 
  & & & \\ 
Impacted by COVID &  & 0.005 & -0.010 \\ 
  &  & (0.007) & (0.035) \\ 
  & & & \\ 
Negative COVID Impact &  & -0.006 & -0.013 \\ 
  &  & (0.007) & (0.032) \\ 
  & & & \\ 
School Skill-set  & 0.122^{***} & 0.006 & -0.077^{**} \\ 
 Matters at Work & (0.013) & (0.006) & (0.034) \\ 
  & & & \\ 
\hline \\[-1.8ex] 
Dep. Mean & 0.24 & 0.17 & 0.37 \\
Sample & \multicolumn{1}{c}{Full} & \multicolumn{1}{c}{Full} & \multicolumn{1}{c}{Still Applying} \\
Observations & \multicolumn{1}{c}{9,568} & \multicolumn{1}{c}{9,568} & \multicolumn{1}{c}{1,687} \\ 
Log Likelihood & \multicolumn{1}{c}{-5,275.892} & \multicolumn{1}{c}{2,004.756} & \multicolumn{1}{c}{-1,175.398} \\ 
Akaike Inf. Crit. & \multicolumn{1}{c}{10,615.780} & \multicolumn{1}{c}{-3,935.511} & \multicolumn{1}{c}{2,422.796} \\ 
\hline 
\hline
\end{tabular}}
\begin{tablenotes}
\item {Note: Each column present selected coefficients of a different Linear Probability Model with the dependent variable specified on top. All specifications include controls for number of friends, gender, age, median income of the home zipcode as well as characteristics of educational opportunities in their high school county. Robust standard errors presented in parentheses. $^{*}$p$<$0.1; $^{**}$p$<$0.05; $^{***}$p$<$0.01}
\end{tablenotes}  
\end{table}

\begin{figure}[hb]
    \centering
    \includegraphics[width=.8\textwidth]{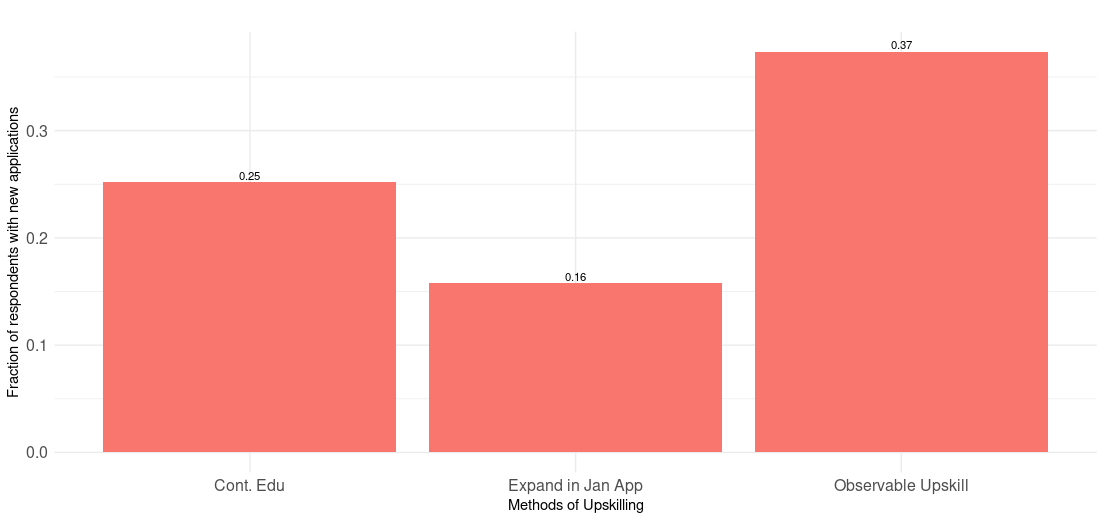}
    \caption{Breakdown of upskilling among users who were still applying to jobs after the survey.}
    \label{fig:upskill_appy}
\end{figure}

We asked users for their latest job title and label these job titles with the most likely ILO skill group associated with it. We also asked users to self-categorize into one of these skill groups. We then label the jobs that users apply to 3 months after the survey to determine whether users were 'observably upskilling.' Figure \ref{fig:upskill_appy} shows the breakdown of upskilling among users who were still applying after our survey. Upskilling through applications, either by 'observably upskilling' through recent applications (37\%) or applications sent in January (16\%) was most common, followed by continued education (25\%). 

To dig deeper, we focus on users still applying to jobs after our survey and in specification 3 of Table 
\ref{tab:upskilling_combined} we model the probability of 'observable upskilling' using a LPM. Users with more education were more likely to send applications to higher-skilled jobs. For example, those with a Bachelor's or Graduate degree were 40\% or 49\% more likely 'observably upskill,' respectively, than users without a degree. Users with a two-year or HS degree were 27\% more likely to apply to higher-skilled jobs than users without a degree. These results are statistically significant at the 10\% level or lower. 


These results are notably different from those on continuing education in that the difference in the likelihood of 'observable-upskilling' by education group is much smaller than the difference in the likelihood of taking coursework by education group. Whereas those with a Bachelor's degree were 37\% more likely to have taken courses than those with a HS degree, they were only 15\% (5.5pp) more likely to have 'observably upskilled' than those with a HS degree -- and the difference is not statistically significant. Given the smaller disparity, initiatives that enable upskilling on the job are likely to have a more equitable impact than initiatives that encourages individuals to take time to learn outside of work.
\FloatBarrier
\section{Conclusion}
\label{sec:conclusion}
Our results highlight the impact of an individual's local network and the local economy on the decision to pursue a higher education, and the perceived return to a higher education. These results indicate that increasing connections between those with different educational backgrounds may influence the decision to pursue an education, especially among younger individuals.  Our data show that 
most users believe skills acquired outside of a formal education are most relevant to their job. Generally speaking, those with a higher degree are more likely to believe that their education has been beneficial for their current job, similar to previous findings \cite{morin2014rising}.



In general, most applicants apply to jobs that are relevant to skills acquired outside of formal education. Those with more education are much more likely to pursue jobs relevant to the skills they acquired through their formal education. While those with more education are more likely to engage in upskilling through both continued education and 'observable upskilling,' where they apply to jobs at a higher skill level than their latest position, the gap is smaller on the 'observable upskilling' dimension. These results suggest that one way to promote economic mobility, and to help those most affected by skill-biased technical change, may be to increase opportunities that provide on-the-job training. Though we cannot speak to this with our data, applications of AR/VR that show promise to train workers in new skills may also be valuable \cite{radianti2020systematic}. While many of these applications have focused on formal classroom learning in STEM fields \cite{villanueva2020meta}, development of these technologies across a broader range of fields or for a workplace setting could fill an important training gap. 

\medskip

\section*{Authors}

\begin{description}
    \item[Monica Bhole] is a Research Scientist at Facebook. She earned a Ph.D. from the Economics Department of Stanford University. 
    \item[Eduardo Laguna Muggenberg] is a Data Scientist at Facebook. He earned a Ph.D. from the Economics Department of Stanford University. 
    \item[Michael Meaney] is a UX Researcher at Facebook. He earned a Ph.D. from the Faculty of Education at the University of Cambridge. 
\end{description}

\medskip

\printbibliography



\end{document}